# Quantum Implementation of Risk Analysis-relevant Copulas

Janusz Milek[1]

(Dated: March 9, 2020)

## Abstract

Modern quantitative risk management relies on an adequate modeling of the tail dependence and a possibly accurate quantification of risk measures, like Value at Risk (VaR), at high confidence levels like 1 in 100 or even 1 in 2000. Quantum computing makes such a quantification quadratically more efficient than the Monte Carlo method; see (Woerner and Egger, 2018) and, for a broader perspective, (Orús *et al.*, 2018). An important element of the risk analysis toolbox is copula, see (Jouanin *et al.*, 2004) regarding financial applications. However, to the best knowledge of the author, no quantum computing implementation for sampling from a risk modeling−relevant copula in *explicit* form has been published so far. Our focus here is implementation of simple yet powerful copula models, capable of a satisfactory capturing the joint tail behaviour of the modelled risk factors. This paper deals with a few simple copula families, including Multivariate B11 (MB11) copula family, presented in (Milek, 2014). We will show that this copula family is suitable for the risk aggregation as it is exceptionally able to reproduce tail dependence structures; see (Embrechts *et al.*, 2016) for a relevant benchmark as well as necessary and sufficient conditions regarding the ultimate feasible bivariate tail dependence structures. It turns out that such a *discretized* copula can be expressed using simple constructs present in the quantum computing: binary fraction expansion format, comonotone/independent random variables, controlled gates, and convex combinations, and is therefore suitable for a quantum computer implementation. This paper presents design behind the quantum implementation circuits, numerical and symbolic simulation results, and experimental validation on IBM quantum computer. The paper proposes also a generic method for quantum implementation of any *discretized* copula.

## I. Introduction

Complex statistical models used in the risk management area are supposed to adequately reproduce relations between risk factors and enable possibly exact quantification of risk, see, *e.g.*, (Jorion, 2007). In particular, an accurate modeling of tail dependence properties is essential for building models of the extreme events, see (Embrechts *et al.*, 2002). In this area models of various risk types are aggregated to quantify *risk-based economic capital,* necessary for financial institutions to be protected against adverse yet probable losses. Following a definition like the one in (McNeil *et al.*, 2015), a *copula* is a multivariate Probability Distribution Function (pdf) with uniform margins on [0−1] support. While *marginal distributions* capture univariate statistical properties of an arbitrary multivariate pdf, these are the *copulas* which fully define dependence amongst the modeled random variables, see (Nguyen and Molinari, 2011) for the application of copulas in *Internal Models* requested by financial regulators. A famous theorem of Sklar shows that these two elements, the copula and marginal distributions, are necessary to fully describe any decent multivariate probability density function, see (Nelsen, 2006) for the details. Additionally, a split between marginal distributions and *explicit* copula may form an alternative to approaches based on modeling of the entire multivariate *pdf*, like in (Zoufal *et. al.*, 2019), and be a natural choice for implementing the widely used copula-based risk aggregation models.

Quantum computing is a rapidly growing field, promising huge speed up in many areas, see (Landsberg, 2018) for a quick introduction or (Nielsen and Chuang, 2010) for the most detailed reference material. Hence it is desired to extend quantum risk analysis toolbox as, *e.g.*, in (Woerner and Egger, 2018) and (Egger *et al.*, 2019), by the copula methods and appropriate copula families.

We need the following definitions:

*Definition 1:* *Upper conditional quantile exceedance probability* between two jointly distributed random variables $X_1$ and $X_2$ (which represent modeled economic losses)
$$cqep_{12}(q) = P(X_1 \geq F^{\leftarrow}_{X_1}(q) | X_2 \geq F^{\leftarrow}_{X_2}(q))$$ where $F^{\leftarrow}_X(q)$ denotes the inverse cumulative probability distribution function of a random variable $X$.

*Definition 2:* *Upper tail dependence* is a limit for $q \to 1^-$ of the above quantity: $$\lambda^u_{12} = \lim_{q \to 1^-} P(X_1 \geq F^{\leftarrow}_{X_1}(q) | X_2 \geq F^{\leftarrow}_{X_2}(q)).$$

Since the 2008 financial crisis the financial market regulators require a satisfactory modeling of the tail dependence, which should be superior to the one of the popular but criticized Gauss copula, see (MacKenzie and Spears, 2014), which asymptotically vanishes, unless the correlation approaches one.

It turns out that a simple yet powerful copula family is *Multivariate B11* (MB11), *i.e.*, multivariate extension of the copula classified by (Joe, 1997) as B11. This copula family has been proposed in (Milek, 2014) for the risk capital aggregation in internal models, however related earlier constructs exist as well, see (Hürlimann, 2002)−for MLS copula, and (Yang *et al.*, 2009) −for the multivariate extension of Fréchet copula. While this copula can be seen as a convex combination of Gauss copulas realizing *extreme correlation matrices,* namely those containing only zero or one entries, it can be calibrated in a very flexible way to reproduce bivariate and multivariate tail dependence structures, see (Milek, 2014) and (Embrechts *et al.*, 2016). It is also interesting to note that, under certain conditions, MB11 copula can reproduce results of the variance-covariance aggregation, popular in the risk management. In case the reader is interested in applying other copula families, note that Appendix E presents a generic method for quantum implementation of any *discretized* copula.

---

1  Independent author, Switzerland. e-mail: janusz.milek@alumni.ethz.ch

## II. MB11 copula family and other copulas

*Definition 3: MB11 copula* is a convex combination (or, mixture) of a number of Gauss copulas, each realizing one particular extreme correlation matrix, containing only zero or one entries. In this note such copulas are called *canonical* copulas. (A decomposition to *canonical* copulas is unique, hence, it is a *canonical* decomposition.) For an *n*-dimensional copula there exists exactly *Bell(n)* canonical copulas, where *Bell* function counts number of ways *a set of n elements can be partitioned into non-empty subsets*, (Bell, 1934). *These subsets represent groups of copula variables, generated using independent uniform random variables* on [0−1], as it is explained below.

### B11 copula

The set {1,2} can be partitioned into non empty subsets in only two distinct ways: {{1,2}} and {{1},{2}}. Hence, the corresponding MB11 copula for *n=2* dimensions becomes a convex sum of two bivariate *fundamental* copulas, the *comonotonicity* $M_2$ and *independence* $\Pi_2$, see (McNeil *et al.*, 2015). Consequently, cumulative *pdf* of the resulting copula can be written as a mixture (convex combination)

$$C_{B11} = \alpha M_2 + (1-\alpha)\Pi_2 \qquad (1)$$

where the non-negative coefficient $\alpha$ represents weight of $M_2$ and non-negative $(1-\alpha)$ weight of $\Pi_2$. Figure 1 depicts probability density function of B11 copula:

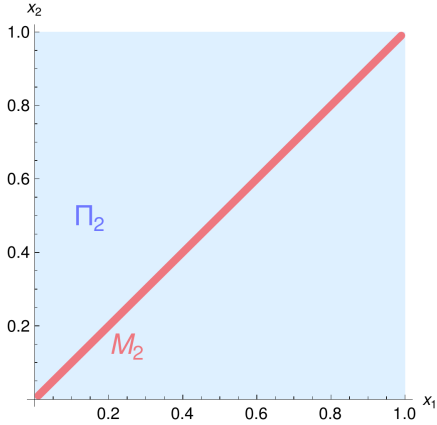

*Figure 1:* Probability density function of B11 copula

The copula variates $x_1$ and $x_2$ can be generated by sampling from the copula mixture (1): variates of $M_2$ consist of replicated independent uniformly distributed random variable on [0−1], while variates of $\Pi_2$ consist of two independent uniformly distributed random variable on [0−1]. This copula is known as B11 copula, as classified by (Joe, 1996) and i*s the bivariate marginal copula* for MB11 copula. It has the following properties regarding dependence measures of its variables $x_1$ and $x_2$: (i) *Spearman's rank correlation coefficient* equals $\alpha$, (ii) *tail dependence coefficient* also equals $\alpha$, (iii) *conditional quantile exceedance probability* is a linear function of probability *p*, decreasing on [0−1] from 1 to $\alpha$:

$$cqep(p) = 1 - p(1-\alpha) \qquad (2)$$

It is remarkable that the tail dependence and correlation coefficients are equal (such that for $\alpha > 0$ the copula ensures a non-zero bivariate tail dependence). This is never the case for the bivariate Gauss copula, unless it is one of the two we use here: $M_2$ or $\Pi_2$.

### Discretization and binary fraction representation

For a quantum computer implementation, each of *n* copula variables, uniformly distributed on [0−1], can be *discretized* to be represented via *k* qubits $|q_1 q_2 ... q_k\rangle$ in the binary fraction expansion format $0.q_1 q_2 ... q_k$, where the dot denotes binary point, see (Bhaskar *et al.*, 2015) for quantum computing data formats.

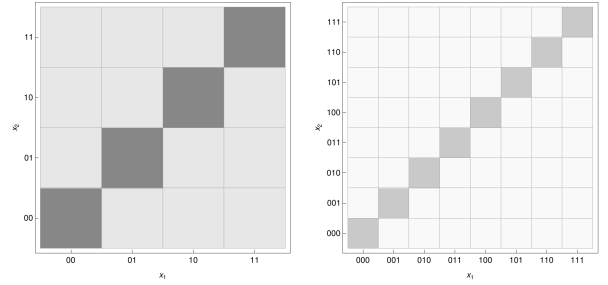

*Figure 2:* Example probability densities of a discretized B11 copula, assuming *k*=2 (left) and 3 (right) qubit resolution per variable

Thus, implementation of an *n* dimensional copula requires *n·k* qubits and discretization of the copula *pdf*; see Appendix E and its equation (25) regarding the discretization process. Figure 2 depicts bivariate *pdf* of B11 copula. Note that density outside the diagonal equals $(1-\alpha)/2^{2k}$ and on the diagonal it increases to $\alpha/2^k + (1-\alpha)/2^{2k}$. A discretized B11 copula preserves both the correlation and conditional quantile exceedance probabilities (2), when evaluated at $\{0, 1, ..., 2^k-1\}/2^k$.

Note that related approaches include checkerboard approximation of (Li *et al.*, 1997) or copula discretization of (Durrleman *et al.*, 2000), see also (Geenens, 2019) for review of copula modelling for discrete random vectors.

### Quantum implementation of fundamental copulas: $M_2$, $\Pi_2$, and $W_2$

Quantum implementation of the three *fundamental* bivariate copulas: *comonotonicity* $M_2$, *independence* $\Pi_2$, and *counter-monotonicity* $W_2$, see (McNeil *et al.*, 2015), is straightforward.

The *comonotonicity* copula $M_2$ with cumulative *pdf*

$$M_2(x_1, x_2) = max(x_1, x_2) \qquad (3)$$

has both copula variables *comonotone*, *i.e.*, $x_2 = x_1$.

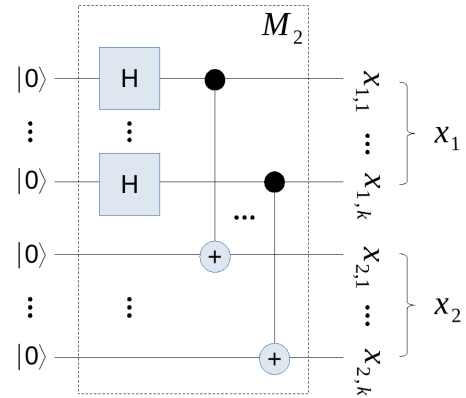

F*igure 3:* Quantum implementation of the bivariate comonotonicity copula $M_2$ with *k* qubit resolution

The first copula variable is generated as $H^{\otimes k}|00...0\rangle_k$ while the second one is a copy of the first one, obtained using *k* CNOT gates. The copula generating circuit is shown in Fig. 3.

The *independence* (or, *product*) copula $\Pi_2$ with cumulative *pdf*

$$\Pi_2(x_1, x_2) = x_1 \cdot x_2 \qquad (4)$$

has both copula variables independent, generated as $H^{\otimes k}|00\ldots 0\rangle_k$. Hence, its formula is $H^{\otimes 2k}|00\ldots 0\rangle_{2k}$ and the corresponding quantum circuit is shown in Fig. 4.

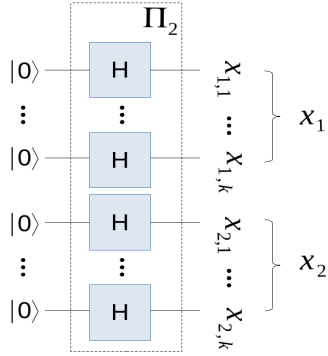

*Figure 4:* Quantum implementation of the bivariate independence copula $\Pi_2$ with *k* qubit resolution per copula variable

The *countermonotonicity* copula $W_2$ with cumulative *pdf*

$$W_2(x_1, x_2) = min(x_1, x_2) \qquad (5)$$

has a pair of *counter-monotone* variables, *i.e.*, $x_2 = 1 - x_1$. The first copula variable can be generated as $H^{\otimes k}|00\ldots 0\rangle$ while the second one obtained as its negated copy, produced using *k* CNOT and X gates, applied to the corresponding qubits. The generating circuit for this copula is shown in Fig. 5.

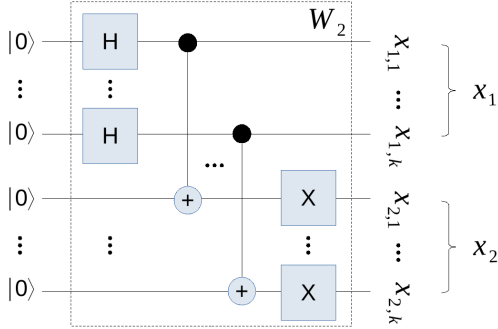

*Figure 5:* Quantum implementation of the bivariate counter-monotonicity copula $W_2$ with *k* qubit resolution per copula variable

## Quantum implementation of B11 copula

The concepts of *pure* and *mixed state* implementation as well as application of control qubits to controlled gates, used in the sequel, are thoroughly presented in (Williams, 2010). First we will review *pure state* implementation options, where the complete state constitutes the copula variates, and thus the number of necessary qubits is the lowest possible. Note that the overall complexity will grow quickly with the qubit resolution.

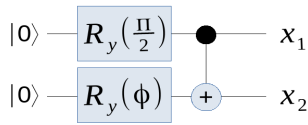

*Figure 6: Pure state* quantum implementation of B11 copula with 1-qubit resolution per copula variable

A *pure state* quantum circuit implementing B11 copula with $k=1-$qubit resolution per variable is shown in Fig. 6, where the angle of the second $R_y$ gate is

$$\phi = 2\arccos\left(\sqrt{\frac{1+\alpha}{2}}\right) \qquad (6)$$

The circuit follows a structure proposed by (Woerner and Egger, 2018) to load discretized probability distribution functions into a quantum computer. It turns out that the above circuit works also for $\alpha<0$, and thus realizes *Linear Spearman* (LS) copula, defined in (Hürlimann, 2002) as

$$C_{LS} = \alpha^+ M_2 + \alpha^- W_2 + (1-|\alpha|)\Pi_2 \qquad (7)$$

It is possible to increase the resolution of B11 copula *pure state* implementation circuit to 2 qubits per copula variable. To design such a circuit it is instrumental to see that the first two qubits, obtained as described above, can be treated not only as *the most significant qubits* of the copula variables, $x_1$ and $x_2$, but also as control qubits to additional bivariate gates producing *the less significant qubits* (this corresponds to *probabilistic conditioning*). These controlled gates are (i) two parallel Hadamard gates, $H^{\otimes 2}$, used to create *independence*, and (ii) another 1-qubit B11 copula circuit, having increased mixing coefficient

$$\alpha_2 = \frac{2\alpha}{1+\alpha} \qquad (8)$$

which determines the angle $\phi_2$ of its rotation gate $R_y$. The corresponding circuit is shown in Fig. 7; note that SWAP gate re-orders the copula qubits to match our notation.

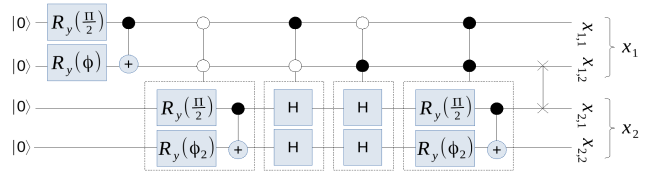

*Figure 7: Pure state* quantum implementation of B11 copula with *k*=2 qubit resolution per copula variable

To handle most general case of *Fréchet* copula, (Nelsen, 2006),

$$C_{Fréchet} = \alpha M_2 + \beta W_2 + (1-\alpha-\beta)\Pi_2 \qquad (9)$$

consider Table 1, which contains state probabilities of the most significant qubits.

|  | $M_2$ | $W_2$ | $\Pi_2$ | Equivalent parameter of B11 copula (*i.e.*, for $\beta=0$) |
|---|---|---|---|---|
| $\|00\rangle$ | $\alpha/2$ | 0 | $(1-\alpha-\beta)/4$ | $\dfrac{2\alpha}{1+\alpha}$ |
| $\|01\rangle$ | 0 | $\beta/2$ | $(1-\alpha-\beta)/4$ | 0 |
| $\|10\rangle$ | 0 | $\beta/2$ | $(1-\alpha-\beta)/4$ | 0 |
| $\|11\rangle$ | $\alpha/2$ | 0 | $(1-\alpha-\beta)/4$ | $\dfrac{2\alpha}{1+\alpha}$ |

*Table 1:* State probabilities of the most significant qubits for bivariate B11 (and Fréchet) copula exhibit mirror symmetry

The less significant qubits of the copula should be generated according to the rows of Table 1, re-normalized to sum to one; note that for generation of further qubits the number of conditioning qubits has to grow. Hence, in the case of the upper copula qubits $|0\ldots 0\rangle_{2k}$ or $|1\ldots 1\rangle_{2k}$ the two *k*-th qubits have to be generated from B11 copula with the dependence coefficient

$$\alpha_k = \frac{2^{k-1}\alpha}{1+(2^{k-1}-1)\alpha} \qquad (10)$$

otherwise *independence* copula $\Pi_2$ has to be used. (Note that as $k$ advances, $\alpha_k$ in (10) asymptotically achieves 1.) We can also extend bivariate B11 copula with one qubit resolution per variable to $n$ variables, it would then become a convex combination of the multivariate independence and comonotonicity copulas, $C = \alpha M_n + (1-\alpha)\Pi_n$ (a subset of MB11 copula). A circuit implementing it is shown in Fig. 8.

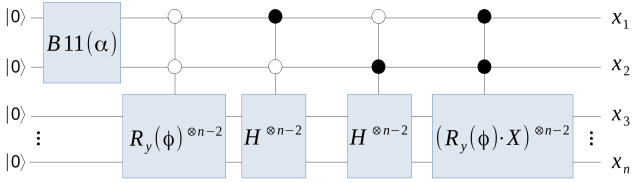

*Figure 8: Pure state* quantum implementation of $n$ dimensional $\alpha M_n + (1-\alpha)\Pi_n$ copula with $k=1$ qubit resolution

The parameter $\phi$ has to be selected as

$$\phi = 2\arcsin\left(\frac{\sqrt{1-\alpha}}{\sqrt{2}\sqrt{1+\alpha}}\right) \qquad (11)$$

A $k$-qubit resolution B11 copula can also be implemented using a more transparent *mixed state* configuration with two controlled subsystems, $M_2$ and $\Pi_2$. Figure 9 depicts this implementation. The upper (not measured, control) qubit is rotated via $R_y$ gate to adjust probability $\alpha$ of being in state |1⟩. In the first (*comonotonicity* $M_2$) subsystem both copula variates are to be identical, therefore qubits of the second variable are copies (produced by $k$ CNOT gates) of those of the first variable (generated using H gates). In the second (*independence* $\Pi_2$) subsystem, all qubits are processed by a set of $2k$ Hadamard gates working in parallel.

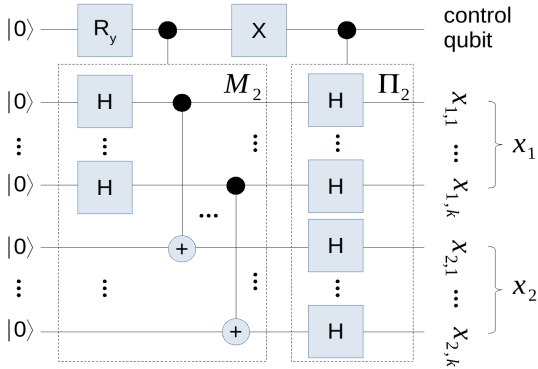

*Figure 9: Mixed state* implementation of B11 copula with one control qubit and $k$ qubit resolution per copula variable

## Multivariate B11 copula in three dimensions

The set {1,2,3} can be partitioned into non empty subsets in five distinct ways: {{1,2,3}}, {{1,2},{3}}, {{1,3},{2}}, {{1}, {2,3}}, and {{1},{2},{3}}. Consequently, the corresponding MB11 copula for $n=3$ dimensions can be written as a *convex combination*

$$C_{MB11} = \alpha_{111}C_{111} + \alpha_{112}C_{112} + \alpha_{121}C_{121} + \alpha_{122}C_{122} + \alpha_{123}C_{123} \qquad (12)$$

where $C_{111}$, $C_{112}$, $C_{121}$, $C_{122}$, and $C_{123}$ are *canonical* copulas, corresponding to the above set partitioning. There, $C_{111}$ is identical to $M_3$, the *comonotonicity* copula in 3 dimensions, $C_{123}$ is identical to $\Pi_3$, the *independence* copula in 3 dimensions, and $C_{112}$, $C_{121}$, and $C_{122}$ resemble *independence* copula $\Pi_2$ with one of the factor variables appearing for two copula variables. The structures of the unitary matrices behind these *canonical* copulas are shown color-coded in Fig. 10-14, as in (Williams, 2010):

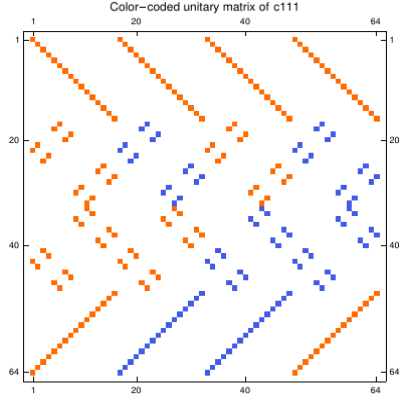

*Figure 10:* The structure of the real unitary matrix behind $C_{111}=M_3$ with $k=2$ qubit resolution per copula variable

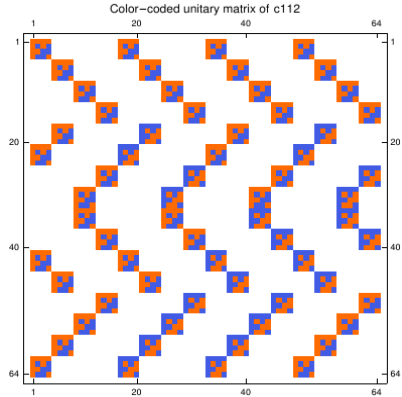

*Figure 11:* The structure of the real unitary matrix behind $C_{112}$ with $k=2$ qubit resolution per copula variable

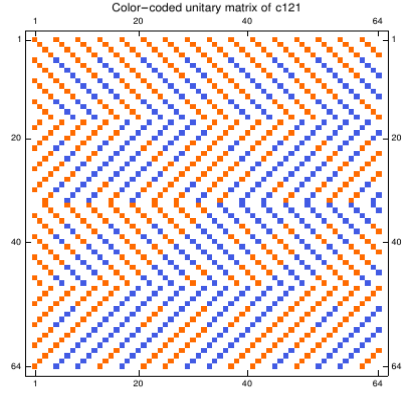

*Figure 12:* The structure of the real unitary matrix behind $C_{121}$ with $k=2$ qubit resolution per copula variable

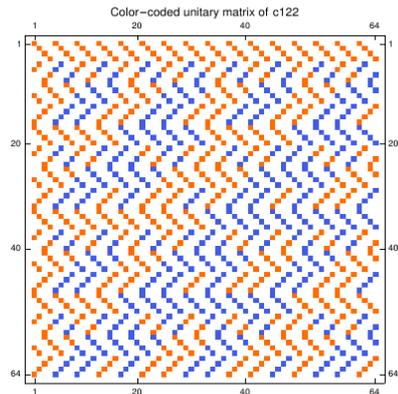

*Figure 13:* The structure of the real unitary matrix behind $C_{122}$ with $k=2$ qubit resolution per copula variable

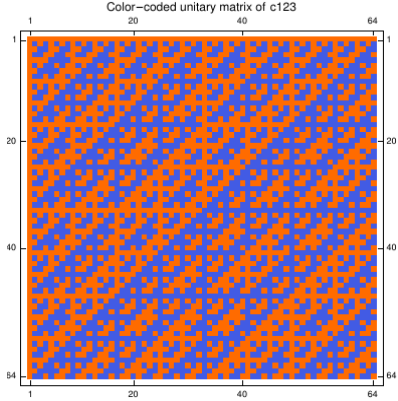

*Figure 14:* The structure of the real unitary matrix behind $C_{123} = \Pi_3$ with $k$=2 qubit resolution per copula variable

The structure of the unitary matrix corresponding to $C_{123}$ in Fig. 14 reveals a kind of the most random pattern as it generates independence. Note that the bivariate tail dependence matrix of the trivariate MB11 copula (12) is (for resolution $k \to \infty$) still identical to Spearman's rank correlation, and equals

$$\Lambda = \begin{bmatrix} 1 & \alpha_{111}+\alpha_{112} & \alpha_{111}+\alpha_{121} \\ \alpha_{111}+\alpha_{112} & 1 & \alpha_{111}+\alpha_{122} \\ \alpha_{111}+\alpha_{121} & \alpha_{111}+\alpha_{122} & 1 \end{bmatrix} \quad (13)$$

what follows from the fact that B11 copula is the marginal bivariate copula of MB11 copula for any $n \geq 3$.

Below we define *upper trivariate tail dependence coefficient*, note that this coefficient is symmetric w.r.t. random variables $X_1$, $X_2$, and $X_3$, see (De Luca and Rivieccio, 2012) for a discussion and examples related to Archimedean copulas.

*Definition 4: Upper trivariate tail dependence* is a limit at $1^-$ of the trivariate conditional quantile exceedance probability, *i.e.*,

$$\lambda_{123}^u = \lim_{q \to 1^-} P(X_1 \geq F_{X_1}^{\leftarrow}(q) \wedge X_2 \geq F_{X_2}^{\leftarrow}(q) | X_3 \geq F_{X_3}^{\leftarrow}(q))$$

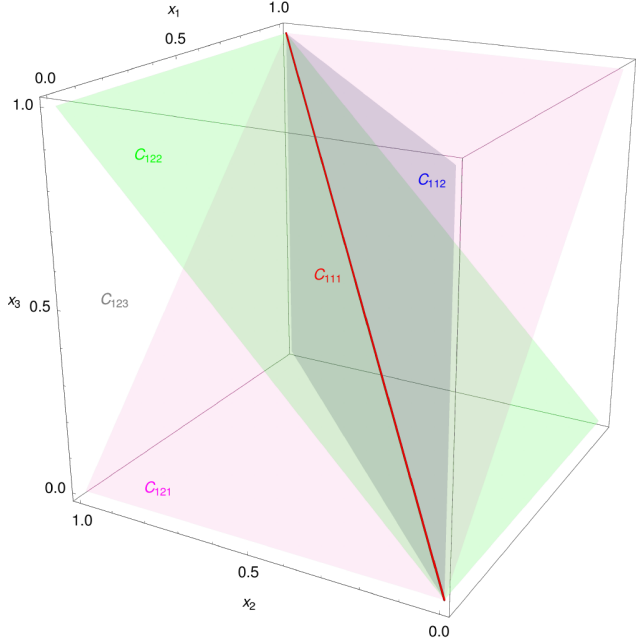

*Figure 15:* Three-dimensional plot of a probability density function for a 3-dimensional MB11 copula

What is relevant for the risk modeling (and the resulting *risk capital*), this copula can ensure a non-zero trivariate tail dependence $\lambda_{123}$, which, to some degree, can be adjusted independently of the other coefficients, and is equal to $\alpha_{111}$. This is a remarkable property, as amongst best known copulas only $t$ copula allows for this, and only if its number of degrees of freedom changes, *i.e.*, at the price of substantially changing also other copula properties. Note that a trivariate MB11 copula is uniquely determined by its bivariate and trivariate tail dependence, as it follows from (15). Example probability density of the copula is shown in Fig. 15. In Figure 16 you will find quantum circuit implementing the trivariate MB11 copula with $k$ qubit resolution per variable. The first three upper control qubits take values in the range $|000\rangle-|100\rangle$ to select one of the five canonical copulas $C_{111}$-$C_{123}$ (see Appendix C for synthesis of the corresponding circuit for *control qubits*). Note that contributions of the canonical copulas to the copula probability density function depend on the copula weights, as shown in Table 2.

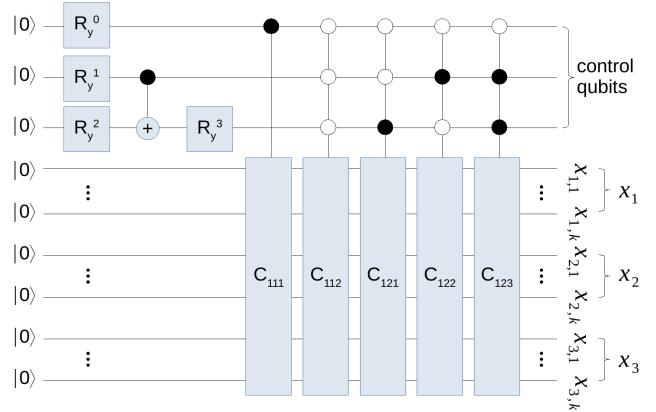

*Figure 16:* Implementation of trivariate MB11 copula with $k$ qubit resolution per copula variable

| copula | $C_{111}$ | $C_{112}$ | $C_{121}$ | $C_{122}$ | $C_{123}$ |
|---|---|---|---|---|---|
| pdf | $\alpha_{111}/2^k$ | $\alpha_{112}/2^{2k}$ | $\alpha_{121}/2^{2k}$ | $\alpha_{122}/2^{2k}$ | $\alpha_{123}/2^{3k}$ |

*Table 2:* Pdf contributions of canonical copulas $C_{111}$-$C_{123}$

## Multivariate B11 copula in *n* dimensions

Generalizing this implementation scheme of trivariate MB11 copula, we propose here a quantum circuit for any number of copula variables, shown in Fig. 17. Note that generation of the control qubits, shown in a draft form in this figure, is addressed, *e.g.*, by (Zoufal *et. al.*, 2019) and some of its references.

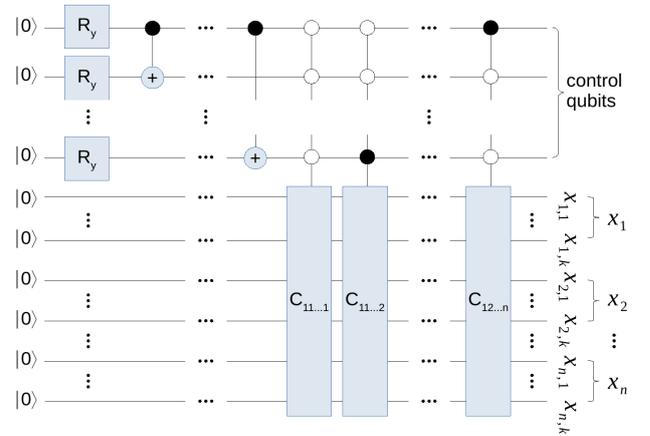

*Figure 17:* Implementation of MB11 copula for arbitrary number of variables and $k$ qubits per copula variable

Note that risk aggregation in *Internal Models* usually involves 6 or more risks, see (Nguyen and Molinari, 2011) for discussion of such models. As the copula dimension $n$ grows, the number of canonical copulas, given by *Bell number* Bn, see (Bell, 1934) or (OEIS, 2020a), increases quickly, as shown in Table 3.

| n  | 2 | 3 | 4  | 5  | 6   | 7   | 8    | 9     | 10     |
|----|---|---|----|----|-----|-----|------|-------|--------|
| Bn | 2 | 5 | 15 | 52 | 203 | 877 | 4140 | 21147 | 115975 |

*Table 3:* Number of canonical copulas given by *Bell number*

This complexity is the source of MB11 copula flexibility, as each of its multivariate tail dependence coefficients can be adjusted, to some degree, separately, see (Milek, 2014). The number of the copula coefficients will, for $n>3$, exceed the number of the tail dependence coefficients. This will lead to an under-determined problem, however, a unique copula parameterization can still be found, *e.g.*, by using convex optimization within *Maximum Entropy* method of (Jaynes, 2003), see (Dai Pra *et al.*, 2013) for a very similar setup (but in a different context, namely of Bell inequalities for quantum states). Another complexity reduction method, proposed in (Milek, 2014) and called *sparse parameterization*, involves multivariate non-linear programming and allows for implementing copulas up to around $n$=75 variables. In the quantum computing realm this method would profit from an application of quantum multiplexers, described, *e.g.*, in (Roy *et al.*, 2012).

## Pure state implementation of trivariate MB11 copula

The *probabilistic conditioning* idea behind the presented earlier pure state implementation of B11 copula and *use of the leading qubits as the control qubits* (see also Appendices D and E) will be applied now to construct a trivariate MB11 copula. Therefore let us begin again with a copula circuit having just $k$=1 qubit resolution per variable (see Appendix A for a discussion). The related state will take $2^3$ values, but $\{0,1\}^3$ cube, containing the corresponding probabilities, will exhibit mirror symmetry in the plane orthogonal to [1,1,1] vector. Such a setup is characterized by 3 degrees of freedom (two identical probability quadruples which sum to ½), while in general trivariate MB11 copula has 4 degrees of freedom (5 coefficients of the convex combination (12) which sum to one). Hence, behaviour of this copula cannot be uniquely determined only via the most significant qubits.

|        | $C_{111}$       | $C_{112}$       | $C_{121}$       | $C_{122}$       | $C_{123}$       |
|--------|-----------------|-----------------|-----------------|-----------------|-----------------|
| $|000\rangle$ | $\alpha_{111}/2$ | $\alpha_{112}/4$ | $\alpha_{121}/4$ | $\alpha_{122}/4$ | $\alpha_{123}/8$ |
| $|001\rangle$ | 0               | $\alpha_{112}/4$ | 0               | 0               | $\alpha_{123}/8$ |
| $|010\rangle$ | 0               | 0               | $\alpha_{121}/4$ | 0               | $\alpha_{123}/8$ |
| $|011\rangle$ | 0               | 0               | 0               | $\alpha_{122}/4$ | $\alpha_{123}/8$ |
| $|100\rangle$ | 0               | 0               | 0               | $\alpha_{122}/4$ | $\alpha_{123}/8$ |
| $|101\rangle$ | 0               | 0               | $\alpha_{121}/4$ | 0               | $\alpha_{123}/8$ |
| $|110\rangle$ | 0               | $\alpha_{112}/4$ | 0               | 0               | $\alpha_{123}/8$ |
| $|111\rangle$ | $\alpha_{111}/2$ | $\alpha_{112}/4$ | $\alpha_{121}/4$ | $\alpha_{122}/4$ | $\alpha_{123}/8$ |

*Table 4:* State probabilities of the most significant qubits for a trivariate MB11 copula exhibit mirror symmetry

In Table 4 we see, *e.g.*, that state $|001\rangle$ for the most significant qubits appears with probability $\alpha_{112}/4 + \alpha_{123}/8$. A quantum circuit in Fig. 18, generating this copula, is essentially a 2−qubit synthesizer for 4 probabilities, extended at the top via one qubit, processed by a Hadamard gate, and controlling two CNOT gates to flip the state of the bottom qubits, see Appendices B and C.

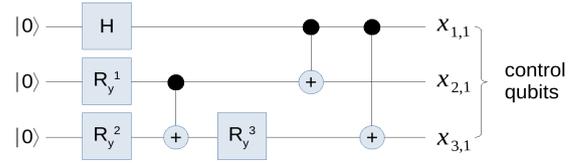

*Figure 18:* Control Qubit Generator CQG for trivariate MB11 copula, generating also the most significant copula qubits

In the next step we add controlled circuits to process additional less significant three qubits, one per copula variable. These circuits are of the same form as for the most significant qubits, but the underlying mixture probabilities of MB11 copula need to be re-scaled to be *probability-compatible* with the control qubits (according to a corresponding row of Table 4) and re-normalized to sum to one. A circuit implementing this approach is shown in Fig. 19.

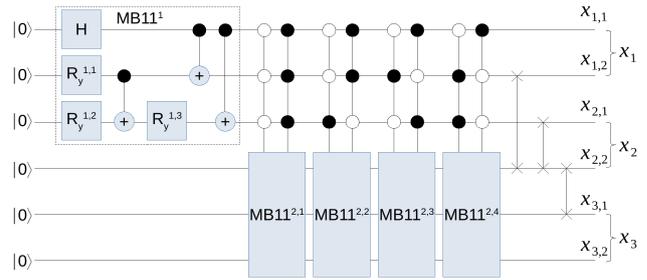

*Figure 19:* Pure state implementation of trivariate MB11 copula with $k$=2 qubit resolution per copula variable

The approach is illustrated using a copula circuit synthesis example assuming the following tail dependence coefficients

$$\begin{aligned}\lambda_{12}&=1/2\\ \lambda_{13}&=1/4\\ \lambda_{23}&=1/8\\ \lambda_{123}&=1/16\end{aligned} \qquad (14)$$

This implies the following probabilities $\alpha_{111}$-$\alpha_{123}$ of the canonical copulas $C_{111}$-$C_{123}$, which have to sum to one:

$$\begin{aligned}\alpha_{111}&=\lambda_{123} & &=1/16\\ \alpha_{112}&=\lambda_{12}-\lambda_{123} & &=7/16\\ \alpha_{121}&=\lambda_{13}-\lambda_{123} & &=3/16\\ \alpha_{122}&=\lambda_{23}-\lambda_{123} & &=1/16\\ \alpha_{123}&=1-\lambda_{12}-\lambda_{13}-\lambda_{23}+2\lambda_{123} & &=1/4\end{aligned} \qquad (15)$$

Note that non-negativity of all these coefficients is equivalent to feasibility of the assumed tail dependence structure. From Table 4 follows that probabilities of states $|000\rangle$-$|011\rangle$ of CQG are

$$\left(\frac{15}{64},\frac{9}{64},\frac{5}{64},\frac{3}{64}\right).$$

For a 2−qubit synthesizer to be used as CQG for 4 probabilities (see (24) in Appendix B) we have the following example solution

$$a_1^1=\frac{\sqrt{3}}{2},\ a_2^1=\frac{1}{2}\sqrt{\frac{1}{2}(4+\sqrt{15})},\ a_3^1=-\frac{1}{\sqrt{2}}$$

automated via Mathematica 12 from (Wolfram Research, Inc., 2020). For the four lower qubit generators we have the following weights of the *canonical copulas*, obtained by the procedure described above

$$\alpha^{2,1}_{111}=\frac{2}{15}, \alpha^{2,1}_{112}=\frac{7}{15}, \alpha^{2,1}_{121}=\frac{1}{5}, \alpha^{2,1}_{122}=\frac{1}{15}, \alpha^{2,1}_{123}=\frac{2}{15}$$

$$\alpha^{2,2}_{111}=0, \alpha^{2,2}_{112}=\frac{7}{9}, \alpha^{2,2}_{121}=0, \alpha^{2,2}_{122}=0, \alpha^{2,2}_{123}=\frac{2}{9}$$

$$\alpha^{2,3}_{111}=0, \alpha^{2,3}_{112}=0, \alpha^{2,3}_{121}=\frac{3}{5}, \alpha^{2,3}_{122}=0, \alpha^{2,3}_{123}=\frac{2}{5}$$

$$\alpha^{2,4}_{111}=0, \alpha^{2,4}_{112}=0, \alpha^{2,4}_{121}=0, \alpha^{2,4}_{122}=\frac{1}{3}, \alpha^{2,4}_{123}=\frac{2}{3}$$

and example (non-unique) solutions of (24):

$$a^{2,1}_1=-\frac{2}{\sqrt{5}},\ a^{2,1}_2=\sqrt{\frac{1}{6}(3-2\sqrt{2})},\ a^{2,1}_3=-\frac{1}{\sqrt{2}}$$

$$a^{2,2}_1=-\frac{2\sqrt{2}}{3},\ a^{2,2}_2=-\frac{1}{\sqrt{2}},\ a^{2,2}_3=-1$$

$$a^{2,3}_1=-\frac{1}{\sqrt{2}},\ a^{2,3}_2=\frac{1}{\sqrt{10}},\ a^{2,3}_3=-\frac{1}{\sqrt{2}}$$

$$a^{2,4}_1=-\frac{1}{\sqrt{2}},\ a^{2,4}_2=-\sqrt{\frac{2}{3}},\ a^{2,4}_3=-1$$

The structure of the unitary matrix behind this implementation is shown in Fig. 20. The above approach can be nested further *top-down* for any number of qubits per copula variable, at the price of the increasing number of the conditioning variables and gates. An example for $k=4$ qubit resolution is provided in Section IV.

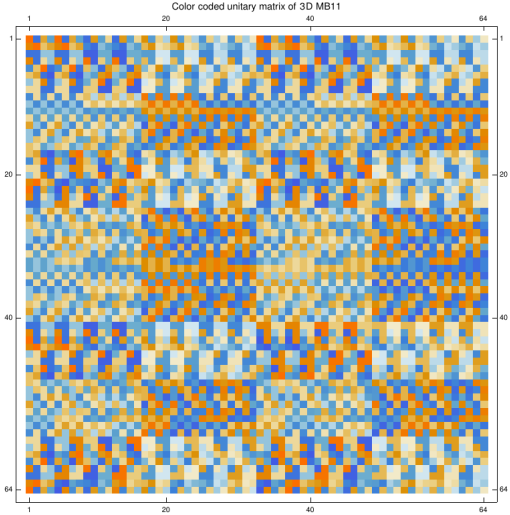

*Figure 20:* The structure of the real unitary matrix behind *pure* state implementation of a trivariate MB11 copula with $k=2$ qubit resolution per variable

## III. Tail dependence benchmarks

Tail dependence benchmarks become a tool to illustrate and test realization of tail dependence structures, see, *e.g.*, (Krause *et al*., 2018) and (Shyamalkumar and Tao, 2019). An initial but ultimate bivariate tail dependence benchmark has been presented by (Milek, 2014). For a four dimensional copula this benchmark requires a copula realization of the bivariate tail dependence matrix (16). Such a matrix cannot be realized neither by t nor by Archimedean copulas, (Hofert, 2012). As shown in (Embrechts *et al.*, 2016), where a necessary and sufficient condition for the *tail dependence compatibility* was provided, this benchmark is sharp, *i.e.*, 1/3 is the upper limit for the constant in the last row and column in this matrix, except the main diagonal.

$$\Lambda_2 = \begin{pmatrix} 1 & 0 & 0 & 1/3 \\ 0 & 1 & 0 & 1/3 \\ 0 & 0 & 1 & 1/3 \\ 1/3 & 1/3 & 1/3 & 1 \end{pmatrix} \qquad (16)$$

It turns out that tail dependence matrix (16) can be realized by the MB11 copula which is the average of $C_{1231}$, $C_{1232}$, and $C_{1233}$ canonical copulas:

$$C = (C_{1231} + C_{1232} + C_{1233}) / 3$$

A quantum computer implementation of such a copula for $k=2$ qubit resolution per variable is shown in Fig. 21, where $R_y$ control gates with the following parameters

$$\phi_1 = 2\arccos\left(\sqrt{\frac{2}{3}}\right)$$
$$\phi_2 = 2\arccos\left(\frac{\sqrt{2+\sqrt{2}}}{2}\right) \qquad (17)$$
$$\phi_3 = 2\arccos\left(-\frac{1}{2}\sqrt{2-\sqrt{2}}\right)$$

produce control states $|00\rangle$, $|01\rangle$, $|10\rangle$, each with probability 1/3.

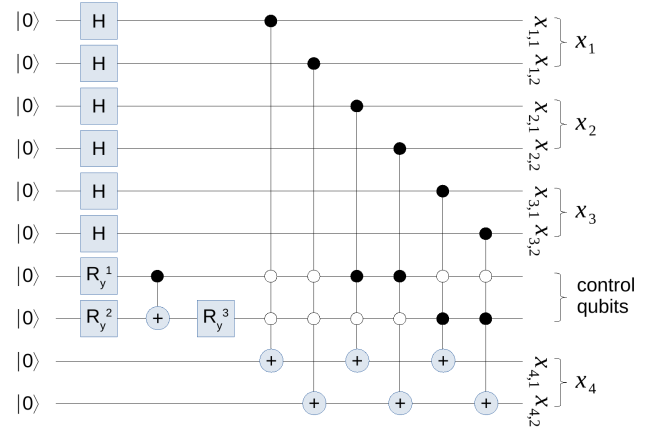

*Figure 21:* 4 dimensional MB11 copula with 2 qubit resolution which corresponds to benchmark (16)

Note that this benchmark can be generalized as follows (while, depending on the parameters, it may loose its sharpness): any bivariate tail dependence structure in *n* dimensions can be realized by MB11 copula, where (i) sum of all entries in tail dependence matrix does not exceed *n*+2, and (ii) individual entries are non-negative.

To prove it, let us observe that the condition is equivalent to requiring that the lower triangular entries of the bivariate tail dependence matrix $\Lambda_2$ are non-negative and sum to at most one. Hence, this MB11 copula must be a convex combination of canonical copulas of the form $C_{123\ldots(n-1)i}$ where the mixing coefficient of $C_{123\ldots(n-1)i}$ copula appearing in the mixture is $\Lambda_2[n,i]$, and the complement of the weights to one is the weight of the *independence* copula, $C_{123\ldots n}$.

## IV. Simulation experiments

This section presents simulation results for some of the copulas presented above. The results were obtained using own symbolic quantum computing modeler, built using Mathematica 12 from (Wolfram Research, Inc., 2020).

## A simple copula with 3 qubit resolution

Let us simulate $C_{MB11} = \frac{1}{2}C_{111} + \frac{1}{2}C_{112}$ copula, assuming three qubit resolution per variable and *mixed state* implementation.

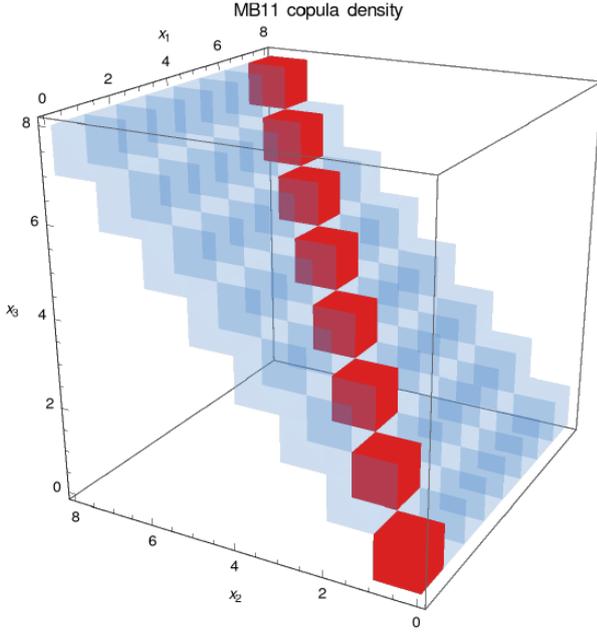

*Figure 22:* 3D plot of a 3-dimensional MB11 copula density with $k=3$ qubit resolution per variable, where the copula is a convex sum of $C_{111}$ and $C_{112}$

The implementation requires 9 qubits for the copula variables and 1 additional control qubit to implement mixture of canonical copulas $C_{111}$ and $C_{112}$. This gives us $2^3=8$ probability levels per variable and 512 discretization cells for $[0-1]^3$ cube. Note that the copula variates are either located on hyperplane $x=y$ or line $x=y=z$, both intersected with $[0-1]^3$ cube. The resulting bivariate correlation (and tail dependence for resolution $k \to \infty$) matrix is

$$\Lambda = \begin{bmatrix} 1 & 1 & 1/2 \\ 1 & 1 & 1/2 \\ 1/2 & 1/2 & 1 \end{bmatrix} \quad (18)$$

while the trivariate tail dependence coefficient (for resolution $k \to \infty$) equals ½ what is the weight of $C_{111}$ copula.

## A more complex copula with 4 qubit resolution

In the second experiment we simulate the previously considered trivariate MB11 copula

$$C_{MB11} = \frac{1}{16}C_{111} + \frac{7}{16}C_{112} + \frac{3}{16}C_{121} + \frac{1}{16}C_{122} + \frac{1}{4}C_{123} \quad (19)$$

with the resolution of four qubits per copula variable. We utilize *pure state* implementation approach, nesting the design from Fig. 19 two more times for the two additional qubits per copula variable. Hence, we now have $4^2=16$ MB11 copula structures in the third and $4^3=64$ in the fourth layer, with the coefficients computed as previously shown for the second layer. This implementation requires only 12 qubits, compared to 15 qubits, necessary for a *mixed state* implementation.

The Spearman's rank correlation (and tail dependence for resolution $k \to \infty$) matrix is

$$\Lambda = \begin{bmatrix} 1 & 1/2 & 1/4 \\ 1/2 & 1 & 1/8 \\ 1/4 & 1/8 & 1 \end{bmatrix} \quad (20)$$

while the trivariate tail dependence coefficient (for $k \to \infty$) is $\alpha_{111}=1/16$. Note that by varying coefficients of the convex combination the most rich set of the bivariate tail dependence structures can be achieved as defined in (Embrechts *et al.*, 2016); this property also holds for $n=4$. The resulting density shown in Fig. 23 becomes a discretized version of the density depicted on Fig. 15 and takes values from Table 2.

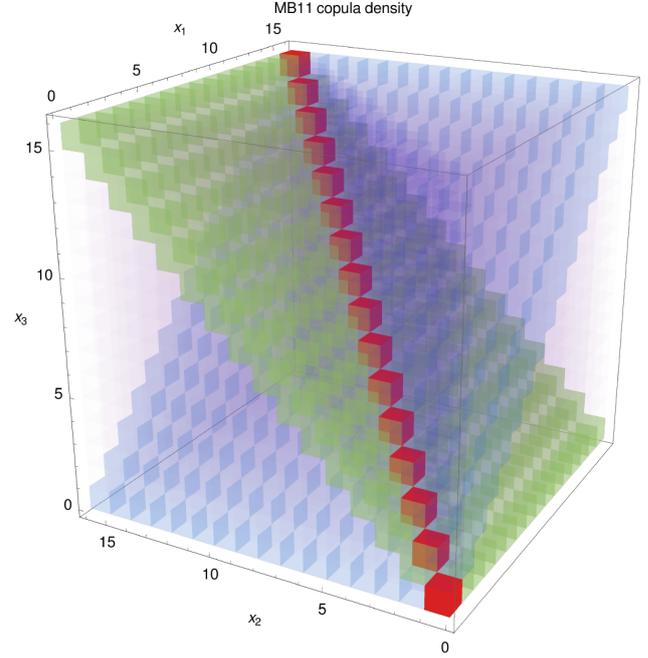

*Figure 23:* Three-dimensional plot of a pdf for a 3-dimensional MB11 copula with $k=4$ qubits per copula variable

Figure 24 presents color coded real unitary matrix behind this implementation.

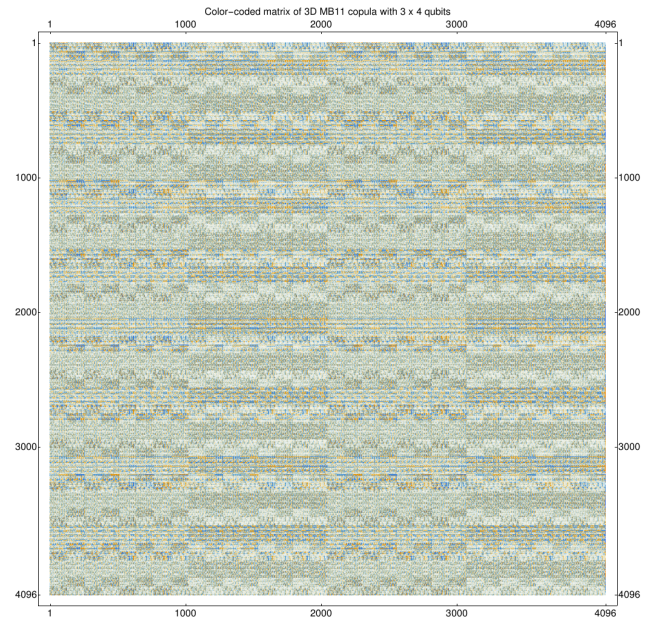

*Figure 24:* The structure of the real unitary matrix behind the trivariate MB11 copula *pure state* generator with $k=4$ qubit resolution per copula variable

# VaR and tail dependence estimation in a B11 copula-based risk model

The presented example follows the framework thoroughly discussed in (Woerner and Egger, 2018) and (Egger *et al.*, 2019). The risk model considered here consists of two risk drivers, $x_1$ and $x_2$:

$$Loss = 16x_1 + 4x_2 \qquad (21)$$

The dependence between the drivers is modeled via B11 copula with $k=2$ qubit resolution per variable and the mixing parameter $\alpha=1/2$. We use the *pure state* implementation from Fig. 7, saving the additional control qubit, otherwise necessary in the *mixed state* implementation. Being copula variables, $x_1$ and $x_2$ individually are uniformly distributed on $\{0,1/4,2/4,3/4\}$ support. Because of the very special weights chosen in (21), the resulting total loss can be directly obtained from the copula variates $.q_1q_2$ and $.q_3q_4$ as $q_1q_2q_3q_4$, and is distributed on $\{0,1,2,…,15\}$ support, as shown in red in Fig. 25. Calculation of Value at Risk, VaR, is performed within a comparator circuit $f(x)=\delta(x \leq v)$, synthesized for each value of the threshold parameter $v$ in the set $\{0,1,...15\}$, as a unitary operation (22):

$$U_{f(.)} = \sum_{x,y} |x\rangle_n \langle x|_n \otimes |y \oplus f(x)\rangle_1 \langle y|_1 \qquad (22)$$

The comparator is realized via *f-controlled-NOT* operation, see (Williams, 2010), p. 43, and uses one additional qubit, denoted $y$, to carry the comparison result. For simplicity, for $x_1$ and $x_2$ we assumed *uniform* marginal distributions on $[0 − 1]$, therefore the copula variates can be directly summed to obtain the total loss (21). However, at least at the mathematical level, it is straightforward to incorporate also any marginal transformation and summation of the losses in this unitary operation. The *estimation of probabilities* necessary to estimate VaR is performed via *bisection* method at the top of the *quantum amplitude and phase estimation* algorithms, see (Woerner and Egger, 2018) and (Egger *et al.*, 2019) for the risk analysis and implementation context, or (Abrams and Lloyd, 1998) and (Brassard *et al.*, 2000) for the properties of the algorithms.

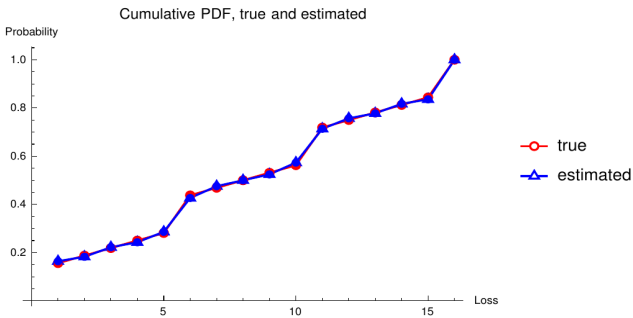

*Figure 25:* True and estimated cumulative pdf of a risk model with MB11 copula dependence

To reduce the number of the necessary qubits, in the synthesis of the implementing circuit we assume to know states $\psi_0$ and $\psi_1$, what could be avoided at the price of using an additional qubit, see (Egger *et al.*, 2019) for the related details. The overall implementation utilizes 12 simulated qubits (4 for copula, 1 for comparator, 7 for probability estimation algorithm). Figure 25 shows true (red) and estimated (blue) VaR values for all the possible loss values. Some discrepancy between true (red) and estimated (blue) values is caused by the discretization of the estimated probability levels to $2^{7-1}=64$ values, as permitted by *probability estimation* algorithm. Note that a substantial speed up of the quantum probability estimation algorithm has been recently proposed in (Grinko *et al.*, 2019).

The *upper conditional quantile exceedance probability* of the copula variables, defined on page 1 as $P(X_1 \geq q \wedge X_2 \geq q) / P(X_1 \geq q)$, can be estimated directly in the quantum simulation circuit also via the aforementioned probability estimation algorithm using $f(x_1,x_2)=\delta(x_1 \geq q)\delta(x_2 \geq q)$ in (22). The algorithm is applied to only estimate $P(X_1 \geq q \wedge X_2 \geq q)$, as from the copula definition it follows $P(X_i \geq q)=1-q$. However, dividing this estimate for higher quantiles by a factor much smaller than one may adversely impact the estimate accuracy, see Fig. 26.

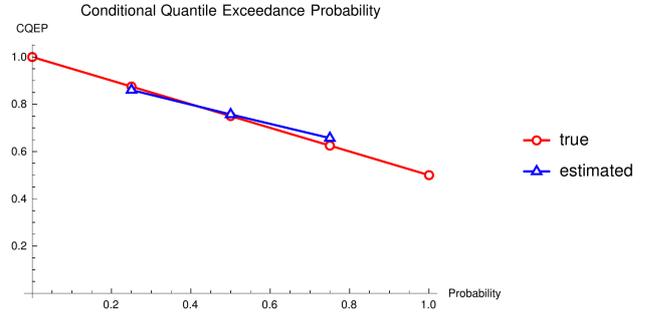

*Figure 26:* Conditional quantile exceedance probability of B11 copula: estimated (blue) and true (red)

# V. Validation on IBM quantum computer

The presented approaches to generation of copula variates are validated, first on IBM QASM simulator, and then IBM quantum computers, for a bivariate B11 copula

$$C_{B11} = \alpha M_2 + (1-\alpha)\Pi_2 \qquad (23)$$

## Pure state implementation, 1 qubit resolution

In this implementation as in Fig. 6 we select $\alpha=1/3$ and $\phi=1.23096$, according to (6). The copula has a one qubit resolution per variable, with the copula variates formed as binary fractions $.q_0$ and $.q_1$. The probability density outside the diagonal is $(1-\alpha)/2^{2k} = 1/6$ and on the diagonal $\alpha/2^k + (1-\alpha)/2^{2k}=1/3$.

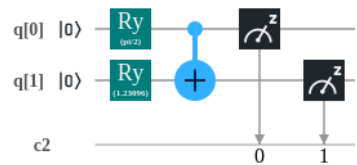

*Figure 27:* Pure state quantum implementation of B11 copula with $k=1$ qubit resolution per variable for $\alpha=1/3$

The results obtained by IBM QASM simulator are shown below:

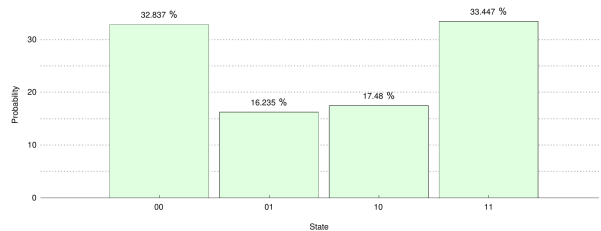

*Figure 28:* Simulation results obtained on IBM QASM simulator

Figure 29 shows results for 5-qubit IBM Vigo quantum computer.

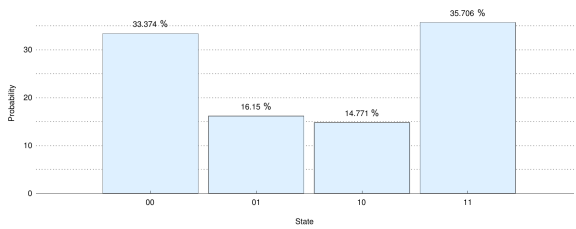

*Figure 29:* Statistics of copula variates generated by 5-qubit IBM Vigo quantum computer for 8192 shots

Note that for this simple circuit there is quite a good match between the theoretic and quantum experiment results.

## Mixed state implementation, 2 qubit resolution

In this implementation we select $\alpha=1/2$. The copula has $k=2$ qubit resolution per variable, with the copula variates formed as binary fractions $.q_0q_3$ and $.q_1q_4$.

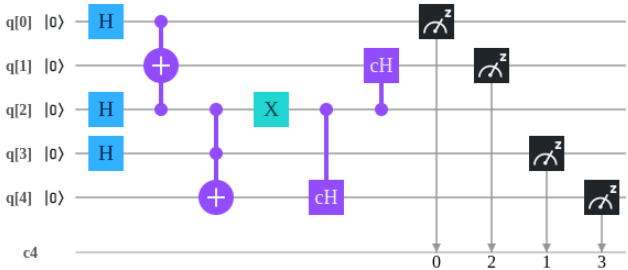

*Figure 30:* Mixed state quantum implementation of B11 copula with $k=2$ qubit resolution per copula variable for $\alpha=1/2$

The control qubit $q_2$ processed by Hadamard gate switches with probability 1/2 between the comonotonicity and independence copulas, such that qubits $q_1$ and $q_4$ are either copies of, correspondingly, $q_0$ and $q_3$, or created independently via the controlled Hadamard gates. The probability density outside the diagonal is $(1-\alpha)/2^{2k}=0.03125$ and on the diagonal $\alpha/2^k + (1-\alpha)/2^{2k}=0.15625$. Example results obtained by IBM QASM simulator are shown in Fig. 31.

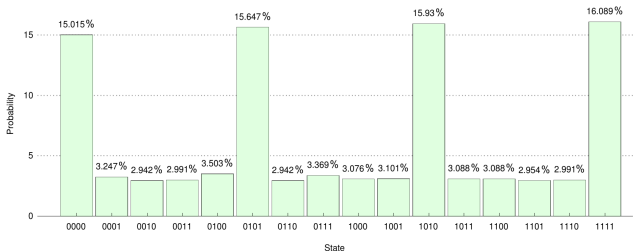

*Figure 31:* Simulation results obtained on IBM QASM simulator

Figure 32 shows example results for 5-qubit IBM QX2 quantum computer.

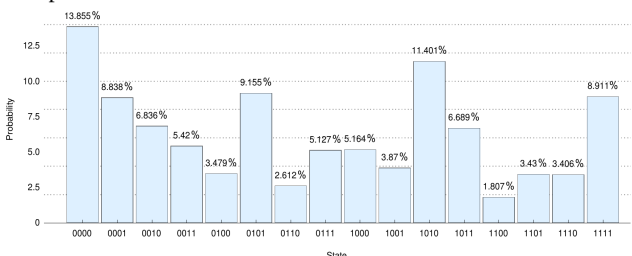

*Figure 32:* Statistics of copula variates generated by 5-qubit IBM QX2 quantum computer for 8192 shots

Comparison of the above results reveals some under-representation of the comonotonicity copula as well as some deviations amongst theoretically equal probabilities of groups of the copula variates.

## VI. Conclusion and outlook

This paper discusses implementation and risk analysis−use of bivariate Fréchet copula, low-dimensional MB11 copula, and other copulas, in the quantum computing context. To author's knowledge, this is the first presentation of an *explicit* risk modeling relevant copula as a quantum circuit.

The proposed approaches can be directly applied to the multivariate extension of Fréchet copula in (Yang *et al.*, 2009); this copula family is interesting because it also covers countermonotonicity, shared, *e.g.*, by *t* copula, recently becoming popular for risk modeling. To achieve it, the *mixed state* implementation of Fig. 17 needs to be extended by additional canonical copulas related to the countermonotone factors, like those for $n=3$ listed in Table 5, and discussed in Appendix D. Alternatively, the pure *state* implementation, as presented for the trivariate MB11 copula, can be utilized, what is also discussed in Appendix D. Note that the number of canonical copulas grows even faster than for MB11, namely as (OEIS, 2020b). But also for this copula family, the *sparse parameterization* approach of (Milek, 2014) would help to reduce the underlying combinatorial complexity.

The quantum computer implementation requires discretization of copulas, but otherwise the results are *exact, i.e.,* without involving approximations. A split into marginal distributions and *explicit* copula (i) may form an alternative to approaches based on modeling of the entire multivariate *pdf*, like in (Zoufal *et. al*., 2019), (ii) constitutes a natural way of implementing the widely used copula-based risk aggregation models, and (3) may bring potential simulation advantages, especially relevant for multi-variate heavy tail statistical models, but this needs to be investigated.

Appendix E presents a generic approach, suitable for quantum implementation of *any* discretized copula. The implementations proposed in the paper are straightforward, as based on first principles, but not optimized. Perhaps, a utilization of specific quantum mechanical effects could bring some complexity reduction.

As a kind of by-product, in Appendix F quantum implementation of a new simple copula class is discussed. The ideas can potentially be of interest for the conventional, non-quantum computing applications, see (Arrazola *et al.*, 2019) for some quantum−inspired algorithms. The paper follows some paradigms of experimental mathematics, see, *e.g.*, (Borwein and Bailey, 2008).

## Acknowledgements

This paper has been created within a personal project. I am very grateful to Gaia Becheri, Federico Degen, and Umut Ordu for suggestions which helped me to prepare and improve it.

Wolfram Mathematica and Mathematica are registered trademarks. IBM and IBM Q Experience are trademarks of International Business Machines Corporation, registered in many jurisdictions worldwide.

The publication reflects only the personal view of the author and does not express nor represent opinions of any particular company nor institution.

# Appendix

## A. Copulas with one qubit resolution

A copula with one qubit resolution per variable represents only a coarse multivariate dependence structure. However, construction of such copulas is simple, since, for each qubit taken individually the states $|0\rangle$ and $|1\rangle$ can be made appearing with equal probabilities, if adjusted by $R_y$ gates.

## B. 2−qubit synthesizer for 4 probabilities

The circuit depicted in Fig. 33 is used in the paper to synthesize two-qubit states with four prescribed probabilities which sum to one. Assumed $\phi_i = 2\arccos(a_i)$ parameterization of $R_y$ gates, counted starting at the upper left corner, leads to the following probabilities of individual states in the computational basis:

$$\begin{aligned}
p_{|00\rangle} &= |a_1 a_2 a_3 - a_1\sqrt{1-a_2^2}\sqrt{1-a_3^2}|^2 \\
p_{|01\rangle} &= |a_1\sqrt{1-a_2^2} a_3 + a_1 a_2\sqrt{1-a_3^2}|^2 \\
p_{|10\rangle} &= |(-1+a_1^2)(\sqrt{1-a_2^2} a_3 - a_2\sqrt{1-a_3^2})|^2 \\
p_{|11\rangle} &= |(-1+a_1^2)(a_2 a_3 + \sqrt{1-a_2^2}\sqrt{1-a_3^2})|^2
\end{aligned} \quad (24)$$

Equations (24) can now be solved for the unknown parameters $a_1$, $a_2$, and $a_3$. The corresponding quantum circuit, if necessary enhanced by a state permutation (requiring X, SWAP, or CNOT gates), can generate arbitrary sets of 4 probabilities which sum to one.

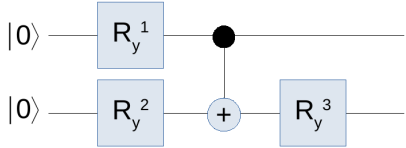

*Figure 33:* Two qubit probability synthesizer

## C. 3−qubit synthesizer for 5 probabilities

The circuit considered here is used in the paper to synthesize three-qubit states with five prescribed probabilities which sum to one.

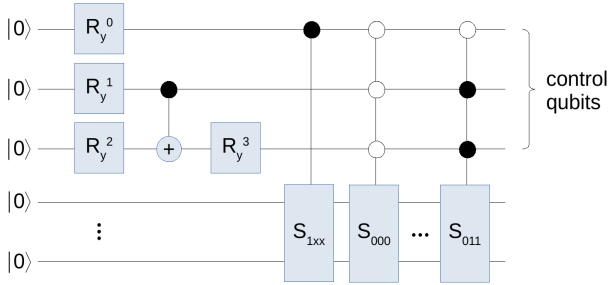

*Figure 34:* Synthesizer for 3 qubits controlling 5 subsystems

These qubits are then used as control qubits in the mixed state implementation of a trivariate MB11 copula. The circuit consists of two qubit probability synthesizer presented above, enhanced by an additional (most significant) qubit, initialized to $|0\rangle$ and processed by an additional $R_y$ gate with parameter $\phi_0 = 2\arccos(a_0)$. The states $|000\rangle$, $|001\rangle$, $|010\rangle$, and $|011\rangle$ will appear with probabilities (24), multiplied by $|a_0|^2$, while the first qubit will be in state $|1\rangle$ with probability $|1-a_0^2|$. Knowing this we can realize arbitrary mixed states requiring application of five controlled subsystems as shown in Fig. 34. Note that the state permutation mentioned in Appendix B is not necessary here, as we can rather re-arrange the controlled subsystems.

## D. Trivariate extension for Fréchet copula

Let us generalize *pure state* implementation of MB11 copula to trivariate Fréchet copula, extending (9) to $n=3$ dimensions. The MB11 partitioning of the set {1,2,3} have to additionally take into account the *countermonotonicity*, marked below using the *minus* sign. Hence, we need to add the following six countermonotone set partitions {{1,-2,3}}, {{1,2,-3}}, {{1,-2,-3}}, {{1,-2},{3}}, {{1,-3},{2}}, and {{1},{2,-3}}. Consequently, Table 4 has to be extended by the additional columns provided in Table 5. The rest of the *pure state* implementation procedure remains the same as for the trivariate MB11 copula, nesting $k-2$ times the structure from Fig. 19.

|  | $C_{1\text{-}11}$ | $C_{11\text{-}1}$ | $C_{1\text{-}1\text{-}1}$ | $C_{1\text{-}12}$ | $C_{12\text{-}1}$ | $C_{12\text{-}2}$ |
|---|---|---|---|---|---|---|
| $|000\rangle$ | 0 | 0 | 0 | 0 | 0 | 0 |
| $|001\rangle$ | 0 | $\alpha_{11\text{-}1}/2$ | 0 | 0 | $\alpha_{12\text{-}1}/4$ | $\alpha_{12\text{-}2}/4$ |
| $|010\rangle$ | $\alpha_{1\text{-}11}/2$ | 0 | 0 | $\alpha_{1\text{-}12}/4$ | 0 | $\alpha_{12\text{-}2}/4$ |
| $|011\rangle$ | 0 | 0 | $\alpha_{1\text{-}1\text{-}1}/2$ | $\alpha_{1\text{-}12}/4$ | $\alpha_{12\text{-}1}/4$ | 0 |
| $|100\rangle$ | 0 | 0 | $\alpha_{1\text{-}1\text{-}1}/2$ | $\alpha_{1\text{-}12}/4$ | $\alpha_{12\text{-}1}/4$ | 0 |
| $|101\rangle$ | $\alpha_{1\text{-}11}/2$ | 0 | 0 | $\alpha_{1\text{-}12}/4$ | 0 | $\alpha_{12\text{-}2}/4$ |
| $|110\rangle$ | 0 | $\alpha_{11\text{-}1}/2$ | 0 | 0 | $\alpha_{12\text{-}1}/4$ | $\alpha_{12\text{-}2}/4$ |
| $|111\rangle$ | 0 | 0 | 0 | 0 | 0 | 0 |

*Table 5:* State probabilities of the remaining canonical copulas of the trivariate extension of Fréchet copula also exhibit mirror symmetry

The resulting example density is shown in Fig. 35 for $k=4$.

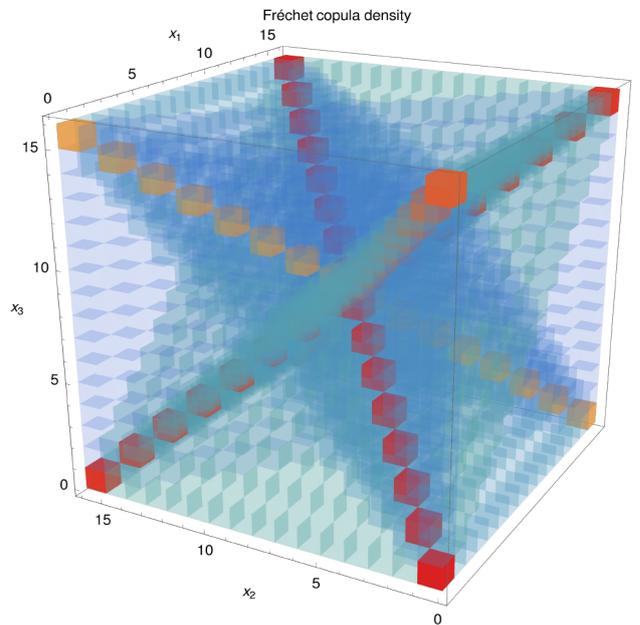

Figure 35: Three-dimensional plot of a pdf for a 3-dimensional Fréchet copula with $k=4$ qubits per copula variable

### E. *Pure state* implementation of any copula

Let us now generalize the quantum *pure state* copula implementation method to *any* copula. The necessary discretized copula probability density function can be computed using its cumulative probability density function $F_{X_{12}}(\cdot)$, evaluated on a suitable discretization grid, and combined using *inclusion-exclusion principle* of Abraham de Moivre, see, *e.g.,* (Stanley, 1986). While an application for the multivariate case is straightforward, for the presentation clarity we consider here only the bivariate copulas, where in two dimensions it holds

$$P(x_1^l \leq x_1 \leq x_1^h \wedge x_2^l \leq x_2 \leq x_2^h) = \\ F_{X_{12}}(x_1^l, x_2^l) - F_{X_{12}}(x_1^l, x_2^h) - F_{X_{12}}(x_1^h, x_2^l) + F_{X_{12}}(x_1^h, x_2^h) \quad (25)$$

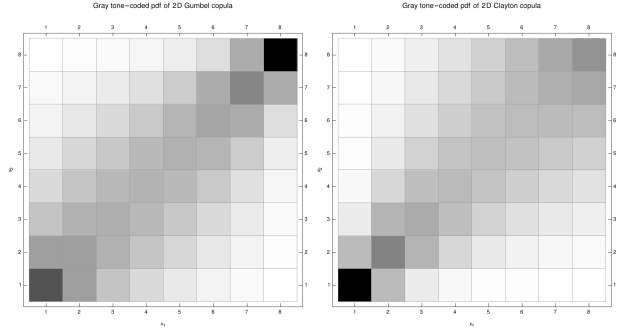

F*igure 36:* Gray tone−coded discretized bivariate Gumbel (left) and Clayton (right) copulas with *k*=3 qubit resolution

As an example, Figure 36 presents probability density functions of Gumbel (left) and Clayton (right) copulas, discretized on $\{0, 1/8, \ldots, 1\}^2$ grid. These copulas exhibit tail dependence (Gumbel−upper, Clayton−lower), and are therefore suitable for a prudent modeling of tail dependent risks, see (McNeil *et al.*, 2015). Also, both belong to Archimedean copula family (Nelsen, 2006), and their cumulative probability density functions are given by

$$F_{Gumbel}(x_1, x_2) = \exp(-((-\ln x_1)^\theta + (-\ln x_2)^\theta)^{1/\theta}), \quad 1 \leq \theta < \infty$$

$$F_{Clayton}(x_1, x_2) = (x_1^{-\theta} + x_2^{-\theta} - 1)^{1/\theta}, \quad 0 < \theta < \infty$$

The proposed approach utilizes a number of 2-*qubit synthesizer* structures, defined in Appendix B and here denoted σ. Each structure, for some integer value *l*, realizes a pair of *l*-th qubits of the two copula variables, and is controlled by (or, conditioned on) values of the upper 1, …, *l*−1 qubits.

The quantum circuit depicted in Fig. 37 should serve as a representative example, as it can realize *any* discretized bivariate copula with *k*=3 qubit resolution per copula variable. The upper structure, $\sigma_1$, generates the first qubits of the copula variables $x_1$ and $x_2$. Generation of the second qubits of the copula variables is conditioned on the values of the first qubits. Probabilities of their occurrence, computed using (25), parametrize synthesizer circuits $\sigma_{2,0}$, $\sigma_{2,1}$, $\sigma_{2,2}$, and $\sigma_{2,3}$. Similarly, generation of the third qubits is conditioned on the first and second qubits of the copula variables, with the synthesizer circuits $\sigma_{3,0,0}, - \sigma_{3,3,3}$.

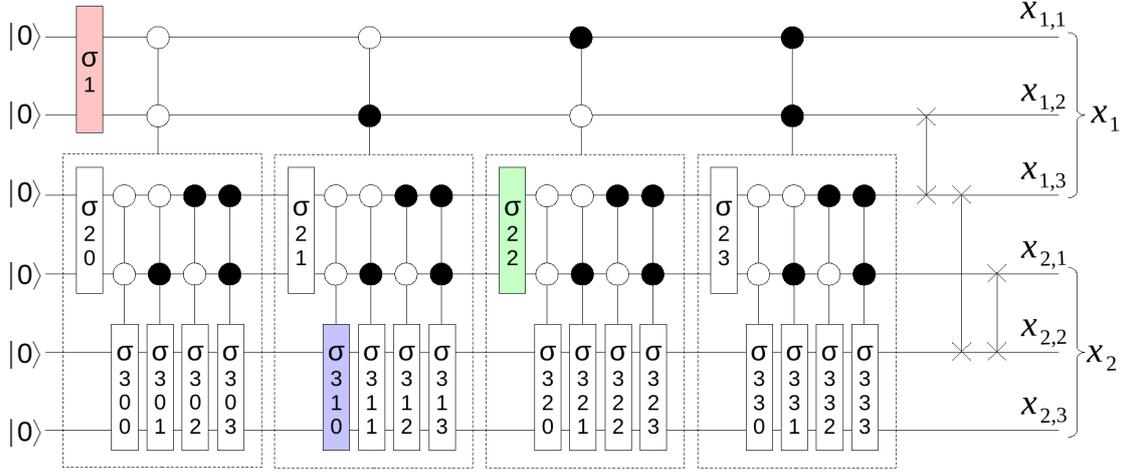

F*igure 37: Pure state i*mplementation of generic bivariate copula with *k*=3 qubit resolution per copula variable

Hence, the conditional probabilities of the qubit states are probabilities of the copula variables taking values in particular quadrants of the discretized pdf's, like the ones shown in Fig. 38 for Gumbel copula, normalized to sum to one.

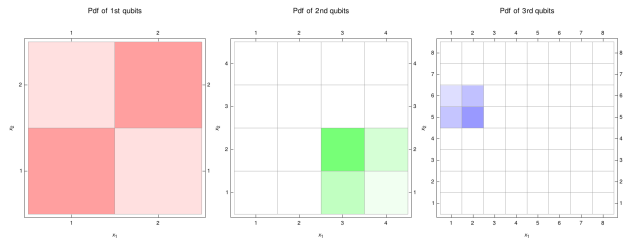

F*igure 38:* Elements of discretized probability density function of Gumbel copula, realized by color-marked gates in Fig. 37

Note that, because of the lack of *radial symmetry* in Gumbel and Clayton copulas, (Nelsen, 2006), in these particular cases the aforementioned probabilities no longer exhibit *mirror symmetry*, utilized in the implementation of MB11 and related copulas, and indicated in Tables 1, 4, and 5.

The presented approach theoretically works for any number of qubits per variable *k* and dimensions *n*. However, the size of the synthesizer structures grows with the copula dimension *n* like $2^n$ and its realization effort increases substantially. Also, the number of synthesizer structures grows exponentially

$$S_k^n = \sum_{l=0}^{k-1} 2^{l \cdot n} = \frac{(2^{k \cdot n} - 1)}{2^n - 1}$$

Hence, the above approach is less efficient than the one for MB11 and Fréchet copulas, since no assumptions about the underlying discretized probability density function are made.

## F. *Copula property* of some quantum circuits

In our search for "quantum copulas", of particular interest are these structures, which, *for any values of their parameters,* constitute a valid copula (*i.e.,* ensure uniform margins on discretized [0−1] support). Clearly, most circuits presented so far in this paper have such a *copula property*. But could we have also other structures? Indeed, one can see that a circuit shown in Fig. 39 represents a copula, since each qubit of a given variable has an equal probability to be measured as zero or one, independently of other qubits of this variable. Due to a resemblance of the generated pattern to cloth material pattern we refer to it as to "fabric" copula. This structure has $k$ parameters (behind $R_y$ gates) but the dependence is defined only at the level of binary digits in the fractional representation of copula variables, therefore this copula may not be necessarily very useful. However, it is a *pure state* copula, therefore it is instructive to check an example probability density function, *e.g.,* in the case when each variable is represented by 4 qubits and its parameters are chosen arbitrarily, as shown in Fig. 40.

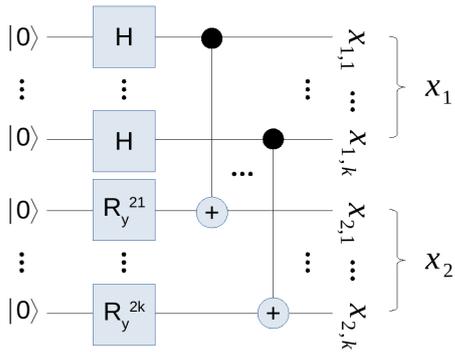

*Figure 39:* Quantum circuit for the bivariate "fabric" copula with $k$ qubit resolution per copula variable

The presented approach can be easily extended to $n$ dimensional copula, as shown in Fig. 41. This copula has $(n-1)$ parameter groups, each containing $k$ parameters $\phi_{jl}$ of $R_y$ gates

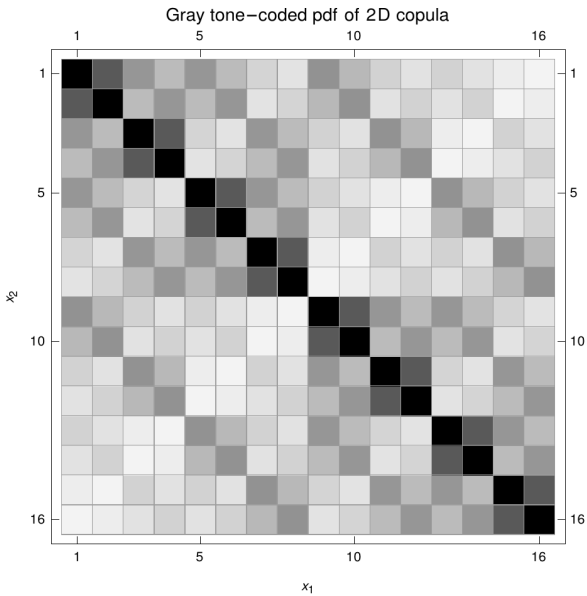

*Figure 40:* Example probability density function of a bivariate "fabric" copula with $k$=4 qubit resolution per copula variable

Again, the parameters in a single group define dependence at the level of the binary digits in the fractional representation of the copula variables. The probability density of each outcome has the form

$$\frac{1}{2^{nk}} \prod_{j,l} (1+(-1)^{f(j,l)} p_{jl}) \qquad (26)$$

where $p_{jl} = \cos(\phi_{jl}/2)^2$, and $f(.)$ is some integer function, but we omit the further details.

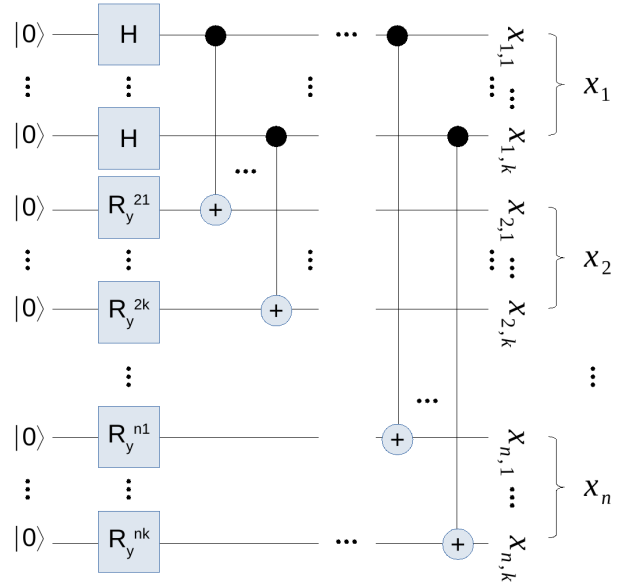

*Figure 41:* Quantum circuit for $n$-variate "fabric" copula with $k$ qubit resolution per copula variable

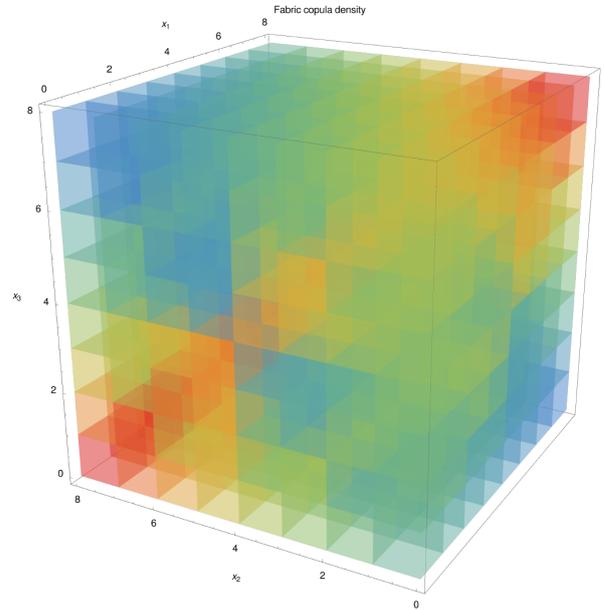

*Figure 42:* Example probability density function of a trivariate "fabric" copula with $k$=3 qubit resolution per copula variable

$$\rho_{12} = \frac{1}{21}(4(p_{11}+p_{21}) - p_{31})$$
$$\rho_{13} = \frac{1}{21}(4(p_{12}+p_{22}) - p_{32}) \qquad (27)$$
$$\rho_{23} = \frac{1}{21}(16 p_{11} p_{21} + 4 p_{12} p_{22} + p_{13} p_{23})$$

Processing formulas for probability density of the bivariate marginal copulas for the trivariate copula with 3 qubit resolution per variable, using Mathematica 12 from (Wolfram Research, Inc., 2020) one can arrive at correlations between its variables given by (27). The first copula variable can treated only as a hidden control variable. In such a case, the correlation between the other copula variables (re-numbered), for the number of qubits growing to infinity will be given by (28).

$$\rho_{ij} = \sum_{k=1}^{\infty} \frac{3}{2^{2k+2}} p_{ik} p_{jk} \qquad (28)$$

Hence, in more than two dimensions, the copula cannot realize all the admissible correlation matrices. Correspondingly, the bivariate tail dependence coefficient becomes

$$\lambda_{ij} = \prod_{k=1}^{\infty} p_{ik} p_{jk} \qquad (29)$$